\documentclass[aps,prl,twocolumn,superscriptaddress,showpacs,floatfix,letterpaper]{revtex4-1}%

\usepackage{graphicx}
\usepackage{bm}
\usepackage{latexsym}
\usepackage{epsf}
\usepackage{rotating}
\usepackage{epsfig,graphics,rotate,color}
\usepackage{wrapfig}
\usepackage{amssymb}
\usepackage{amsmath}
\usepackage{amsfonts}
\usepackage{subfigure}
\usepackage{array,hhline,dcolumn}%
\usepackage[normalem]{ulem}
\usepackage{color}
\usepackage{color}

\providecommand{\U}[1]{\protect\rule{.1in}{.1in}}

\def\ta{\tau}

\def\ch{\chi}

\def\om{\omega}

\def\De{\Delta}

\def\nuebar{\bar\nu_e}
\def\numubar{\bar\nu_\mu}
\def\nutaubar{\bar\nu_\tau}

\newcommand{\beq}{\begin{eqnarray}}
\newcommand{\eeq}{\end{eqnarray}}
\def\to{\rightarrow}

\def\no{\nonumber}
\def\fr#1#2{\frac{#1}{#2}}

\def\aL{(a_L)}
\def\cL{(c_L)}
\def\As#1{({\cal A}_s)_{#1}}
\def\Ac#1{({\cal A}_c)_{#1}}
\def\Bs#1{({\cal B}_s)_{#1}}
\def\Bc#1{({\cal B}_c)_{#1}}
\def\C#1{({\cal C})_{#1}}
\def\indxmb{{\bar e\bar\mu}}
\def\indxtb{{\bar e\bar\tau}}
\def\indxn{{e\mu}}

\def\indxt{{e \ta}}

\def\uGeV{\mbox{GeV}}
\def\uMeV{\mbox{MeV}}

\def\liveday{227.9}
\def\power{4.25}
\def\baseline{1050}
\def\IBDcan{8249}
\def\bkgd{497}
\def\Enu{4.2}

\def\DCOscPp{91.8}

\def\dchipf{60.0}
\def\dchipt{41.8}

\def\fkopf{70.1}\def\fktpf{94.5}
\def\fkopt{68.8}\def\fktpt{94.7}

\def\chot{3.5}\def\chof{5.9}
\def\chtt{8.0}\def\chtf{11.3}

\def\RCffv{ 5.8}\def\RCffl{7.8}
\def\Asffv{-0.4}\def\Asffl{6.6}
\def\Acffv{ 0.4}\def\Acffl{7.0}
\def\Bsffv{ 0.0}\def\Bsffl{5.4}
\def\Bcffv{ 0.5}\def\Bcffl{5.4}

\def\RCtfv{ 5.8}\def\RCtfe{1.7}
\def\Astfv{-0.4}\def\Astfe{0.7}\def\Astfl{1.9}
\def\Actfv{ 0.5}\def\Actfe{0.8}\def\Actfl{5.5}

\def\RCsfv{34.2}\def\RCsfe{9.2}

\mathchardef\mhyphen="2D

\begin{document}

\title{First Test of Lorentz Violation with a Reactor-based Antineutrino Experiment}
\newcommand{\Aachen}{III. Physikalisches Institut, RWTH Aachen University, 52056 Aachen, Germany}
\newcommand{\Alabama}{Department of Physics and Astronomy, University of Alabama, Tuscaloosa, Alabama 35487, USA}
\newcommand{\Argonne}{Argonne National Laboratory, Argonne, Illinois 60439, USA}
\newcommand{\APC}{APC, AstroParticule et Cosmologie, Universit\'{e} Paris Diderot, CNRS/IN2P3, CEA/IRFU, Observatoire de Paris, Sorbonne Paris Cit\'{e}, 75205 Paris Cedex 13, France}
\newcommand{\CBPF}{Centro Brasileiro de Pesquisas F\'{i}sicas, Rio de Janeiro, RJ, cep 22290-180, Brazil}
\newcommand{\Chicago}{The Enrico Fermi Institute, The University of Chicago, Chicago, IL 60637, USA}
\newcommand{\CIEMAT}{Centro de Investigaciones Energ\'{e}ticas, Medioambientales y Tecnol\'{o}gicas, CIEMAT, E-28040, Madrid, Spain}
\newcommand{\Columbia}{Columbia University; New York, NY 10027, USA}
\newcommand{\Davis}{University of California, Davis, CA-95616-8677, USA}
\newcommand{\Drexel}{Physics Department, Drexel University, Philadelphia, Pennsylvania 19104, USA}
\newcommand{\Hamburg}{Institut f\"{u}r Experimentalphysik, Universit\"{a}t Hamburg, 22761 Hamburg, Germany}
\newcommand{\Hiroshima}{Hiroshima Institute of Technology, Hiroshima, 731-5193, Japan}
\newcommand{\IIT}{Department of Physics, Illinois Institute of Technology, Chicago, Illinois 60616, USA}
\newcommand{\INR}{Institute of Nuclear Research of the Russian Aacademy of Science, Russia}
\newcommand{\CEA}{Commissariat \`{a} l'Energie Atomique et aux Energies Alternatives, Centre de Saclay, IRFU, 91191 Gif-sur-Yvette, France}
\newcommand{\Livermore}{Lawrence Livermore National Laboratory, Livermore, CA 94550, USA}
\newcommand{\Kansas}{Department of Physics, Kansas State University, Manhattan, Kansas 66506, USA}
\newcommand{\Kobe}{Department of Physics, Kobe University, Kobe, 657-8501, Japan}
\newcommand{\Kurchatov}{NRC Kurchatov Institute, 123182 Moscow, Russia}
\newcommand{\MIT}{Massachusetts Institute of Technology; Cambridge, MA 02139, USA}
\newcommand{\MaxPlanck}{Max-Planck-Institut f\"{u}r Kernphysik, 69117 Heidelberg, Germany}
\newcommand{\Niigata}{Department of Physics, Niigata University, Niigata, 950-2181, Japan}
\newcommand{\NotreDame}{University of Notre Dame, Notre Dame, IN 46556-5670, USA}
\newcommand{\IPHC}{IPHC, Universit\'{e} de Strasbourg, CNRS/IN2P3, F-67037 Strasbourg, France}
\newcommand{\SUBATECH}{SUBATECH, CNRS/IN2P3, Universit\'{e} de Nantes, Ecole des Mines de Nantes, F-44307 Nantes, France}
\newcommand{\Sussex}{Department of Physics and Astronomy, University of Sussex, Falmer, Brighton BN1 9QH, United Kingdom}
\newcommand{\Tennessee}{Department of Physics and Astronomy, University of Tennessee, Knoxville, Tennessee 37996, USA}
\newcommand{\TokyoInst}{Department of Physics, Tokyo Institute of Technology, Tokyo, 152-8551, Japan  }
\newcommand{\TokyoMet}{Department of Physics, Tokyo Metropolitan University, Tokyo, 192-0397, Japan}
\newcommand{\TohokuUni}{Research Center for Neutrino Science, Tohoku University, Sendai 980-8578, Japan}
\newcommand{\Muenchen}{Physik Department, Technische Universit\"{a}t M\"{u}nchen, 85747 Garching, Germany}
\newcommand{\TohokuGakuin}{Tohoku Gakuin University, Sendai, 981-3193, Japan}
\newcommand{\Tubingen}{Kepler Center for Astro and Particle Physics, Universit\"{a}t T\"{u}bingen, 72076, T\"{u}bingen, Germany}
\newcommand{\UFABC}{Universidade Federal do ABC, UFABC, Sao Paulo, Santo Andr\'{e}, SP, Brazil}
\newcommand{\UNICAMP}{Universidade Estadual de Campinas-UNICAMP, Campinas, SP, Brazil}
\newcommand{\Aviette}{Laboratoire Neutrino de Champagne Ardenne, domaine d'Aviette, 08600 Rancennes, France}
\newcommand{\VirginiaTech}{Center for Neutrino Physics, Virginia Tech, Blacksburg, Virginia 24061, USA}
\newcommand{\deceased}{Deceased.}

\affiliation{\Aachen}
\affiliation{\Alabama}
\affiliation{\Argonne}
\affiliation{\APC}
\affiliation{\CBPF}
\affiliation{\Chicago}
\affiliation{\CIEMAT}
\affiliation{\Columbia}
\affiliation{\Davis}
\affiliation{\Drexel}
\affiliation{\Hamburg}
\affiliation{\Hiroshima}
\affiliation{\IIT}
\affiliation{\INR}
\affiliation{\CEA}
\affiliation{\Livermore}
\affiliation{\Kansas}
\affiliation{\Kobe}
\affiliation{\Kurchatov}
\affiliation{\MIT}
\affiliation{\MaxPlanck}
\affiliation{\Niigata}
\affiliation{\NotreDame}
\affiliation{\IPHC}
\affiliation{\SUBATECH}
\affiliation{\Sussex}
\affiliation{\Tennessee}
\affiliation{\TokyoInst}
\affiliation{\TokyoMet}
\affiliation{\TohokuUni}
\affiliation{\Muenchen}
\affiliation{\TohokuGakuin}
\affiliation{\Tubingen}
\affiliation{\UFABC}
\affiliation{\UNICAMP}

\author{Y.~Abe}
\affiliation{\TokyoInst}

\author{C.~Aberle}
\affiliation{\MaxPlanck}


\author{J.C.~dos Anjos}
\affiliation{\CBPF}




\author{M.~Bergevin}
\affiliation{\Davis}

\author{A.~Bernstein}
\affiliation{\Livermore}

\author{T.J.C.~Bezerra}
\affiliation{\TohokuUni}

\author{L.~Bezrukhov}
\affiliation{\INR}

\author{E.~Blucher}
\affiliation{\Chicago}


\author{N.S.~Bowden}
\affiliation{\Livermore}

\author{C.~Buck}
\affiliation{\MaxPlanck}

% ...
\collaboration{Double Chooz Collaboration}
\begin{abstract} 
We present a search for Lorentz violation with 8249 candidate electron antineutrino events taken by the Double Chooz experiment in 227.9 live days of running. This analysis, featuring a search for a sidereal time dependence of the events, is the first test of Lorentz invariance using a reactor-based antineutrino source. No sidereal variation is present in the data and the disappearance results are consistent with sidereal time independent oscillations. Under the Standard-Model Extension (SME), we set the first limits on fourteen Lorentz violating coefficients associated with transitions between electron and tau flavor, and set two competitive limits associated with transitions between electron and muon flavor. 
\end{abstract}

\maketitle

Recently, we reported evidence of electron antineutrino disappearance with the Double Chooz far detector, $\baseline$~m away from two $\power$~GW reactor cores~\cite{dc1stpub,300daypaper}, which generally is interpreted in terms of mass-induced neutrino oscillations. A path to more exotic physics beyond the Standard Model (SM) may be gained by carefully examining the oscillation behavior. In particular, the collected electron antineutrino sample also provides an opportunity to search for the violation of Lorentz invariance. 

Lorentz invariance requires that the behavior of a particle is independent of its direction or boost velocity. The as-yet-unseen violation of this principle is predicted to occur at the Planck scale and is especially interesting as it can occur dynamically via spontaneous Lorentz symmetry breaking~\cite{SLSB}. The process of neutrino oscillation, in which a neutrino of one flavor transforms into another flavor after traveling a distance, is due to interference between the slightly different Hamiltonian eigenstates of the propagating particle. The experimental observable, oscillation probability, is therefore quite sensitive to small couplings between neutrinos and a possible Lorentz violating field.  

Testing Lorentz violation with the natural interferometer of neutrino oscillation has been done in several experiments, including MINOS~\cite{minos_lv1,*minos_lv2, *minos_lv3}, IceCube~\cite{icecube_lv}, LSND~\cite{lsnd_lv}, and MiniBooNE~\cite{miniboone_lv}. These tests all fall under the formalism of a search within the Standard-Model Extension (SME)~\cite{SME1,*SME2,*SME3}. In this paper, we describe the first search for Lorentz violation using reactor antineutrinos.

In the SME, all possible types of Lorentz violation are 
added to the SM Lagrangian. Here, we limit ourselves to the renormalizable sector (referred to as the minimal SME). For Lorentz violating neutrino oscillation, the effective Hamiltonian is written as~\cite{hamiltonian}
\begin{equation}
(h^\nu_{\rm{eff}})_{ab}\sim
\frac{(m^2)_{ab}}{2E}+
\frac{1}{E}[(a_L)^{\mu} p_{\mu}-(c_L)^{\mu\nu}p_{\mu}p_{\nu}]_{ab}~,
\label{eq:hamiltonian} 
\end{equation}
where $E$ and $p_{\mu}$ are the energy and 4-momentum of the neutrino and $(m^2)_{ab}$ refers to the neutrino mass in the flavor basis represented by $a$ and $b$. 
The CPT-odd coefficient $\aL^{\mu}_{ab}$ switches sign for antineutrinos and violates both Lorentz and CPT symmetry, 
while the CPT-even coefficient $\cL^{\mu\nu}_{ab}$ violates Lorentz but maintains CPT. Both are vector and tensor and consist of direction independent parts 
[$\aL^T_{ab}$, $\aL^Z_{ab}$, $\cL^{TT}_{ab}$, $\cL^{TZ}_{ab}$, and $\cL^{ZZ}_{ab}$] 
and direction dependent parts [$\aL^X_{ab}$, $\aL^Y_{ab}$, $\cL^{TX}_{ab}$, $\cL^{TY}_{ab}$, $\cL^{XX}_{ab}$, $\cL^{XY}_{ab}$, $\cL^{XZ}_{ab}$, $\cL^{YY}_{ab}$, and $\cL^{YZ}_{ab}$] in the Sun-centered coordinate system (represented by the superscripts). A measured non-zero direction dependent component would be clear evidence of an anisotropy in the Universe and Lorentz violation. As discussed later, no evidence for Lorentz violation has been found and our goal is therefore to set limits on these coefficients. We note that the known neutrino mass term in the flavor basis in this formalism is neglected in order to follow a conservative approach when setting these limits. 

Sidereal time is based on the Earth's orientation relative to the fixed stars. The unambiguous signature of Lorentz violation is a sidereal modulation of an experimental observable such as neutrino oscillation probability. A sidereal variation is expected for an experiment moving in a fixed Lorentz violating field with the rotation of the Earth. We probe this field by searching for such a dependence among the collected electron antineutrino events. The antineutrino vector is set using the antineutrino source and the location of the detector. The location of the source is taken to be a weighted point in between the two cores, 6$^\circ$ apart relative to the detector, representative of the number of antineutrinos expected from each during the physics run.

The data set used for this analysis was obtained with the Double Chooz
experiment between April 13, 2011 and March 15, 2012. Electron antineutrinos interact in the detector via the inverse beta decay (IBD) process, $\bar \nu_e + p \rightarrow e^+ + n$. IBD events produce a distinct double coincidence signature from the prompt positron signal followed by neutron capture $30~\mu \mathrm{s}$ (mean time) later. The ``inner detector," composed of three concentric cylindrical regions separated by acrylic, is used to observe and efficiently reconstruct these events as well as mitigate background. The innermost 10~m$^3$ cylinder contains 1~g/l gadolinium-doped scintillator and forms the antineutrino target. Surrounding this is the ``gamma catcher," designed to detect gamma rays escaping the target volume. The gamma catcher volume is then enveloped by a non-scintillating oil buffer in which 390~10~inch PMTs are immersed. The inner detector is surrounded by a steel vessel that forms an optically isolated outer cylinder filled with scintillator. This ``inner veto'', along with an ``outer veto" mounted above it and 15~cm of shielding steel, is used to reject cosmic ray events.  

The antineutrino sample and event selection criteria are identical to those used for the disappearance analysis reported in Ref.~\cite{300daypaper}. The data consist of $\IBDcan$~IBD candidates, collected with $\liveday$~live days and 33.7~GW-ton-years exposure. There are $\bkgd$~background events expected in this sample. The background is mainly composed of (1) cosmogenic radioisotopes, such as $^8$He and $^9$Li, which decay via the emission of $\beta n$, (2) cosmogenic stopping muons as well as fast neutrons that interact multiple times in the inner detector, and (3) accidental coincidence of a radioactivity-induced prompt signal followed by a neutron-like signal. Background event rate as a function of sidereal time is treated as a constant. As the dominant background contributions to the Double Chooz analyses arise as the result of cosmic ray muons, we study the time dependence of muon veto rate in order to justify this assumption. The maximum variation in muon veto rate as a function of sidereal time is about 0.5\%. A background variation in time at this level would create a maximum variation in disappearance probability of $\sim$0.03\%. 

\begin{figure}[ht]
\begin{center}
\mbox{
\epsfxsize=0.52\textwidth 
\hspace{-.3cm}
\epsfbox{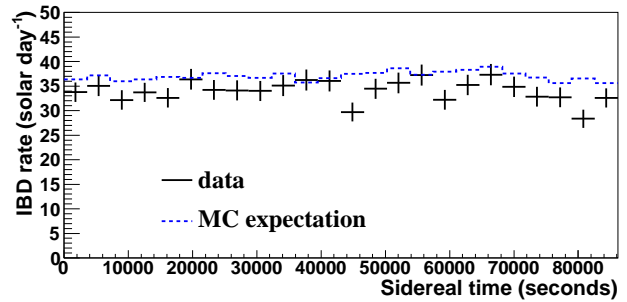}
}
\vspace{-1cm}
\end{center}
\caption{The background subtracted data and MC expectation IBD event rates as a function of sidereal time. The MC expectation assumes no antineutrino disappearance. Total errors (statistical and systematic) are shown on the data points.
}
\label{fig:dataMC}
\end{figure}

The background-subtracted IBD candidate sample is directly compared to the Monte Carlo (MC) expectation in order to probe a possible sidereal time dependence. The unoscillated MC expectation is based on the IBD cross section, the reactor flux prediction, the detector response, and the number of protons in the detection volume. The expectation is formed from each of these variables on a run-by-run basis, with each physics run lasting approximately one hour. We note that the thermal power of each core is estimated in $<$1~minute time intervals and the uncertainty on the total power is 0.5\%. The reactor flux prediction uses extensive input from the Chooz reactor facility and \'{E}lectricit\'{e} de France (EDF). The quality of the code has been benchmarked~\cite{Takahama} and compared to EDF assembly simulations. The $\bar \nu_e$ spectrum is taken directly from Refs.~\cite{huber,Mueller} and is normalized to the Bugey4 rate measurement~\cite{Bugey4}. The analysis input information, shown in Fig.~\ref{fig:dataMC}, is assigned to one of 24 bins between 0 and 23.934 hours (one sidereal day). A MC expectation event weight is split up between the relevant time bins based on the time and length of the run while a data event is placed in a bin based on its DAQ time stamp. 

A number of sources of systematic uncertainty are considered. These include those associated with the background prediction, 
the detector and detector response, and the reactor flux (normalization and shape). The reactor flux and detector operations are both weak functions of solar time due to human activity as the cores turned on/off multiple times during the run and detector calibrations are generally done during the daylight hours. Day-night effects are well accounted for in the MC prediction. All uncertainties are included in a covariance matrix, fully describing the predicted statistical and systematic errors. The 3.93~minute/day difference between sidereal and solar time, compounded over the $\sim$1~year physics run, largely removes any potential for an unaccounted modulation in sidereal time associated with small modulations related to solar dependence. The detector and background prediction uncertainties are considered uncorrelated with each other and fully correlated in sidereal time. A thorough explanation of the various uncertainties and their determination can be found in Ref.~\cite{300daypaper}, noting that correlations in time (as opposed to antineutrino energy in the reference) are most important here. The total fractional uncertainty with respect to the MC expectation is 2.9\%. The statistical uncertainty contributes at the level of 1.1\% and systematic uncertainties are led by the reactor flux and detector response (1.7\%) and the background prediction (1.7\%).

In the three active flavor neutrino oscillation framework, 
the $\nuebar\to\nuebar$ probability can be written as a function of $\nuebar\to\numubar$ and $\nuebar\to\nutaubar$ ($P_{\nuebar\to\nuebar}=1-P_{\nuebar\to\numubar}$$-$$P_{\nuebar\to\nutaubar}$). Under the SME, both $P_{\nuebar\to\numubar}$ and $P_{\nuebar\to\nutaubar}$ are written as functions of five free parameters~\cite{SBA}:
\beq
&P&_{\nuebar\to\nuebar}\simeq1-\frac{|(h_{\mathrm{eff}})_{\indxmb}|^2 L^2}{(\hbar c)^2}
 -\frac{|(h_{\mathrm{eff}})_{\indxtb}|^2 L^2}{(\hbar c)^2}\no\\
&=&1-\fr{L^2}{(\hbar c)^2} [\,|\,\C\indxmb 
  +\As\indxmb \sin \om_\oplus T_\oplus +\Ac\indxmb \cos \om_\oplus T_\oplus  \no\\
&+&  \Bs\indxmb \sin2\om_\oplus T_\oplus \, 
+\Bc\indxmb \cos2\om_\oplus T_\oplus |^2 \no\\
  &+& |\,\C\indxtb 
+ \As\indxtb \sin \om_\oplus T_\oplus 
+\Ac\indxtb \cos \om_\oplus T_\oplus \no\\ 
&+&\Bs\indxtb \sin2\om_\oplus T_\oplus 
+\Bc\indxtb \cos2\om_\oplus T_\oplus \,|^2\,] \label{eq:SBAfull}
\eeq
The disappearance probability is a function of sidereal time~$T_\oplus$, 
sidereal frequency~$\om_\oplus$ [$2\pi/86164.1~\mathrm{\frac{rad}{s}}$], 
baseline $L$, and ten amplitudes (parameters). 
The parameters themselves are composed of the Lorentz violating coefficients introduced in Eq.~\ref{eq:hamiltonian}, antineutrino energy, and the antineutrino-source-to-detector vector. We aim to reduce this equation since there are too many parameters for a realistic fit and measurement extraction. Ideally, this reduction proceeds without any assumptions, in a model independent way.

Double Chooz's maximum sensitivity to the CPT-odd and CPT-even SME coefficients is on the order of $\sim$$10^{-20}$~GeV and $\sim$$10^{-18}$, respectively, determined by considering the maximum oscillation condition in the effective Hamiltonian (Eq.~\ref{eq:hamiltonian}). Noting that the effective Hamiltonian is Hermitian and $\mu-e$ results can be applied to $e-\mu$, the MINOS near detector~\cite{minos_lv1} and MiniBooNE~\cite{miniboone_lv} measurements both place significantly better limits on all CPT-even coefficients, at the level of $\sim$$10^{-21}$ and $\sim$$10^{-20}$, respectively. The ten relevant SME coefficients are therefore set to zero, corresponding to the removal of two parameters, $\Bs\indxmb$ and $\Bc\indxmb$, from Eq.~\ref{eq:SBAfull}. It is now difficult to remove more parameters in a model independent way and we cannot reduce Eq.~\ref{eq:SBAfull} further using existing measurements. We therefore study two different sets of assumptions.

The assumption that all Lorentz violating oscillations occur in electron antineutrino to tau antineutrino transitions ($P_{\nuebar\to\numubar}=0$, $P_{\nuebar\to\nutaubar}\neq 0$) is studied with the ``$e-\ta$ fit":  
\beq P_{\nuebar\to\nuebar}
&\simeq&1-\fr{L^2}{(\hbar c)^2} [\,|\,\C\indxtb
+\As\indxtb \sin \om_\oplus T_\oplus \no\\
&&+\Ac\indxtb \cos \om_\oplus T_\oplus
+\Bs\indxtb \sin2\om_\oplus T_\oplus \no\\
&&+\Bc\indxtb \cos2\om_\oplus T_\oplus \,|^2\,]\label{eq:SBA5} 
\eeq
The five free parameters [$\C\indxtb$, $\As\indxtb$, $\Ac\indxtb$, $\Bs\indxtb$, and $\Bc\indxtb$] themselves contain fourteen of the $e-\tau$ sector SME coefficients introduced in Eq.~\ref{eq:hamiltonian}. 

The second model is based on the assumption that all Lorentz violating oscillations occur in electron antineutrino to muon antineutrino transitions
($P_{\nuebar\to\numubar}\neq 0$, $P_{\nuebar\to\nutaubar}=0$) and is referred to as the ``$e-\mu$ fit": 
\beq
P_{\nuebar\to\nuebar}
&\simeq&1-\fr{L^2}{(\hbar c)^2} [\,|\,\C\indxmb
  +\As\indxmb \sin \om_\oplus T_\oplus \no\\
&&+\Ac\indxmb \cos \om_\oplus T_\oplus \,|^2\,]\label{eq:SBA3}
\eeq
This equation has only three free parameters [$\C\indxmb$, $\As\indxmb$, and $\Ac\indxmb$], as the MINOS and MiniBooNE constraints have removed the CPT-even coefficients, and contains four $e-\mu$ sector SME coefficients. The $\cal{C}$ parameter in Eqs.~\ref{eq:SBA5} and \ref{eq:SBA3} contains sidereal time independent SME coefficients. This term can affect both shape and normalization. We note that each of these two models considers disappearance in only one channel while the complete formula (Eq.~\ref{eq:SBAfull}) contains contributions from both. However, the limits reported tend to be more conservative than if both channels were considered simultaneously.

The SME parameters in Eqs.~\ref{eq:SBA5} and \ref{eq:SBA3} are extracted using the MC expectation and background subtracted data (Fig.~\ref{fig:dataMC}), the total error matrix including statistical and correlated systematic contributions, and a least squares fitting technique. The least squares estimator is minimized in order to find the best fit (BF) among the parameter combinations.

\begin{figure}[ht]
\begin{center}
\mbox{
\epsfxsize=0.52\textwidth 
\hspace{-.3cm}
\epsfbox{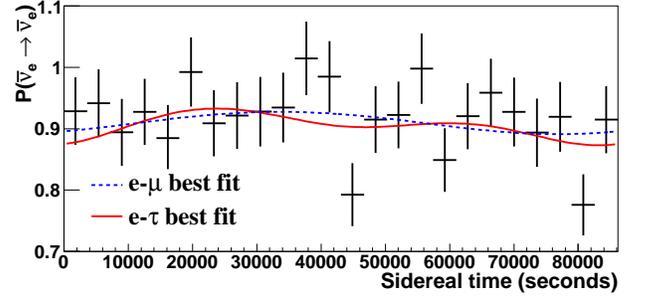}
}
\vspace{-1cm}
\end{center}
\caption{The electron antineutrino disappearance probability as a function of sidereal time, overlaid with the best fit $e-\mu$ ($\chi^2/ndf=28.8/21$) and $e-\ta$ ($\chi^2/ndf=27.7/19$) curves. A one parameter flat fit to the distribution yields a best fit with $\chi^2/ndf=30.6/23$.
}
\label{fig:ratio}
\end{figure}

The $e-\ta$ and $e-\mu$ BF sidereal time results are shown in Fig.~\ref{fig:ratio}. The BF results for both fits are dominated by the sidereal time independent terms, $\C\indxtb$ and $\C\indxmb$. We examine the significance of the results below.

The flatness of the sidereal time distribution is analyzed using a frequentist approach. A large sample of randomized pseudoexperiments, based on the MC expectation and the total error matrix and with an injected sidereal time independent (``flat") disappearance, is generated in order to determine the fraction of samples that present a more or less flat solution than the one found here. We introduce a normalization factor of $\DCOscPp$\%, consistent with the counting-only disappearance probability, in order to ensure that we are testing the null hypothesis that there is no sidereal time dependence rather than the null hypothesis that there is no antineutrino disappearance. The $\De\ch^2$ is defined as the minimum $\ch^2$ from the flat hypothesis minus the minimum $\ch^2$ from each $e-\ta$ or $e-\mu$ fit. This frequentist study shows that $\dchipf$\% ($\dchipt$\%) of pseudoexperiments have a larger $\De\ch^2$ than the real data and that the $e-\ta$ ($e-\mu$) results are consistent with sidereal time independent oscillations. In the absence of a sidereal dependence, we proceed to set limits on the relevant time dependent SME coefficients.
\begin{table*}[!ht]
\begin{center}
\footnotesize
\begin{tabular}{lccc}
\hline
\hline
&~~~BF parameter~~~~&Upper limit&SME coefficients combination \\
&($10^{-20}$~GeV)&(95\%~CL)& \\
\hline
$\C\indxtb$ &$\RCffv$&$\RCffl$&
[$-\aL^T_\indxt-0.29\aL^Z_\indxt]+E[-1.46\cL^{TT}_\indxt-0.57\cL^{TZ}_\indxt+0.38\cL^{ZZ}_\indxt]$\\
$\As\indxtb$&$\Asffv$&$\Asffl$&
$[-0.91\aL^X_\indxt+0.29\aL^Y_\indxt]+E[-1.83\cL^{TX}_\indxt+0.58\cL^{TY}_\indxt-0.52\cL^{XZ}_\indxt+0.16\cL^{YZ}_\indxt]$\\
$\Ac\indxtb$&$\Acffv$&$\Acffl$&
$[ 0.29\aL^X_\indxt+0.91\aL^Y_\indxt]+E[ 0.58\cL^{TX}_\indxt+1.83\cL^{TY}_\indxt+0.16\cL^{XZ}_\indxt+0.52\cL^{YZ}_\indxt]$\\
$\Bs\indxtb$&$\Bsffv$&$\Bsffl$&
$E[ 0.26(\cL^{XX}_\indxt-\cL^{YY}_\indxt)+0.75\cL^{XY}_\indxt]$\\ 
$\Bc\indxtb$&$\Bcffv$&$\Bcffl$&
$E[ 0.38(\cL^{XX}_\indxt-\cL^{YY}_\indxt)-0.53\cL^{XY}_\indxt]$\\ \hline
$\C\indxmb$ &$\RCtfv\pm\RCtfe$& --- &$[-\aL^T_\indxn-0.29\aL^Z_\indxn]$\\
$\As\indxmb$&$\Astfv\pm\Astfe$&$\Astfl$&
$[-0.91\aL^X_\indxn+0.29\aL^Y_\indxn]$\\
$\Ac\indxmb$&$\Actfv\pm\Actfe$&$\Actfl$&
$[ 0.29\aL^X_\indxn+0.91\aL^Y_\indxn]$\\ \hline
\hline
\end{tabular}
\end{center}
\caption{
A summary of the $e-\ta$ and $e-\mu$ Lorentz violation measurements in terms of the best fit (BF) parameters and the corresponding combinations of Standard-Model Extension coefficients. The allowed regions and limits reported are set by the extremes of the multi-dimensional confidence regions. The average antineutrino energy ``$E$'' is $4.2\times10^{-3}$~GeV. }
\label{tab:SME}
\end{table*}
Limits on the SME coefficients and allowed regions around the BF parameters are determined by constructing a five and three dimensional parameter space, corresponding to the $e-\ta$ and $e-\mu$ fits, respectively. By assuming the minimum of the least squares fit estimator follows a $\ch^2$ distribution, a 68\%~CL (95\%~CL) hyper-volume can be defined as the region enclosed by 
the constant $\ch^2$ hyper-surface with minimum $\ch^2$ plus $\chof$ ($\chtf$) for the $e-\ta$ fit, and $\chot$ ($\chtt$) for the $e-\mu$ fit. These criteria are tested by using a sample of pseudoexperiments with an injected signal based on the BF. That is, each pseudoexperiment sample is convolved with the BF oscillation probability equation. A new fit is then performed and the BF parameters are tallied. We find that the above choices for 68\%~CL (95\%~CL) hyper-surfaces enclose $\fkopf$\% ($\fkopt$\%) and $\fktpf$\% ($\fktpt$\%) of BF points for the $e-\ta$ ($e-\mu$) fits and that our allowed regions are valid. Note that we have considered only half of the parameter space in this procedure and that the sign reversed BF parameters are equally valid.

The results are summarized in Table~\ref{tab:SME}. The BF values from both the $e-\ta$ and $e-\mu$ fits are shown along with 68\%~CL allowed regions and 95\%~CL upper limits, when applicable.  The allowed regions are generally asymmetric; however, the larger of the two-sided region is reported. Correlations between parameters and multiple connected solutions make it impossible to extract meaningful allowed regions for the $e-\ta$ fit. The combination of SME coefficients associated with each measured parameter is also shown in the table. All $e-\ta$ parameter limits as well as the sidereal time dependent $e-\mu$ limits are on the order of $\sim$$10^{-20}~\uGeV$ for CPT-odd coefficients and $\sim$$10^{-17}$ for CPT-even coefficients. 

Although every measured sidereal time dependent parameter is consistent with zero, 
the time independent parameter $\C\indxmb$ is non-zero at the 96\%~CL. We note that a normalization-only fit [$P_{\nuebar\to\nuebar}\simeq 1-\fr{L^2}{(\hbar c)^2}( \C\indxtb^2+\C\indxmb^2)$] yields $\C\indxtb^2+\C\indxmb^2=(\RCsfv\pm\RCsfe)\times 10^{-40}~\mathrm{GeV}^2$. This disappearance is consistent with the rate-only $\theta_{13}$ measurement in Ref.~\cite{300daypaper}. With current precision, time independent Lorentz violating effects cannot be distinguished from mass and $\theta_{13}$ induced oscillations. Separating the two effects may be possible with future high statistics data and spectral information, however. The disappearance observed can generally be interpreted as due to neutrino mass and $\theta_{13}$ in the three flavor neutrino oscillation framework.

There are a number of alternative neutrino oscillation models 
motivated by Lorentz violation~\cite{KM2,Tandem,BMW,Puma1,Puma2}. 
These models neglect sidereal modulations by assuming that any such variations 
are averaged out or that the probability of oscillation is governed by time independent terms only. The models focus on reproducing the global observed energy and baseline dependence of neutrino oscillations. Interestingly, however, none of the models predict the observed antineutrino oscillations at Double Chooz's energy ($\langle E \rangle$=$\Enu~\uMeV$) and baseline ($\baseline$~m). That is, the measured disappearance conflicts with these models. This may be an additional reason to interpret the time independent disappearance observed as due to neutrino mass and non-zero $\theta_{13}$, rather than time independent Lorentz violation.

We have analyzed the sidereal time dependence of Double Chooz's electron antineutrino candidates as a probe of Lorentz violation. With no observed modulation, we set the first limits on fourteen of the SME coefficients in the $e-\ta$ sector, and set competitive limits on two $e-\mu$ sector coefficients. Competitive limits may also be provided by other reactor antineutrino experiments in the future~\cite{dayabay,reno}, although Double Chooz features a comparatively simple antineutrino-source-to-detector vector. With the addition of this work amongst the world's data, sidereal variation tests with neutrino oscillation experiments have been performed with all active oscillation channels. In the future, astrophysical neutrinos~\cite{icecube_uhe} may improve sensitivity to Lorentz violation by many orders of magnitude compared to what is possible for terrestrial neutrino experiments.

We thank the French electricity company EDF; the
European fund FEDER; the R\'egion de Champagne Ardenne; 
the D\'epartement des Ardennes; and the Communaut\'e des Communes
Ardennes Rives de Meuse. We acknowledge
the support of the CEA, CNRS/IN2P3, the computer center CCIN2P3, and 
LabEx UnivEarthS in France; the Ministry of Education, Culture,
Sports, Science and Technology of Japan (MEXT) and
the Japan Society for the Promotion of Science (JSPS);
the Department of Energy and the National Science
Foundation of the United States; the Ministerio de Ciencia
e Innovaci´on (MICINN) of Spain; the Max Planck
Gesellschaft, and the Deutsche Forschungsgemeinschaft
DFG (SBH WI 2152), the Transregional Collaborative and Structure of the Universe, and the Maier-Leibnitz-
Laboratorium Garching in Germany; the Russian Academy of Science,
the Kurchatov Institute and RFBR (the Russian
Foundation for Basic Research); the Brazilian Ministry
of Science, Technology and Innovation (MCTI), the Financiadora
de Estudos e Projetos (FINEP), the Conselho Nacional de Desenvolvimento Cient\'{i}fico e Tecnol\'{o}gico 
(CNPq), the S\~{a}o Paulo Research Foundation (FAPESP), and the Brazilian Network for High Energy Physics (RENAFAE) in Brazil. We also thank Alan Kosteleck\'{y} for valuable discussions.
\bibliography{LV_PRDbib}
\end{document}